# On the reducibility of geometric constraint graphs

Samy Ait-Aoudia[1], Adel Moussaoui[1], Khaled Abid[1], Dominique Michelucci[2]
[1]ESI – Ecole nationale Supérieure en Informatique (ex INI), Oued-Smar, Alger, Algérie
s_ait_aoudia@esi.dz, k_abid@esi.dz, a_moussaoui@esi.dz
[2]Laboratoire LE2I, Université de Bourgogne, Dijon, France
dominique.michelucci@u-bourgogne.fr

**Abstract.**

Geometric modeling by constraints, whose applications are of interest to communities from various fields such as mechanical engineering, computer aided design, symbolic computation or molecular chemistry, is now integrated into standard modeling tools. In this discipline, a geometric form is specified by the relations that the components of this form must verify instead of explicitly specifying these components. The purpose of the resolution is to deduce the form satisfying all these constraints. Various methods have been proposed to solve this problem. We will focus on the so-called graph-based or graph-based methods with application to the two-dimensional space.

## I. INTRODUCTION

Geometry modeling by constraints allows users to specify geometric objects such as points, lines, and circles by constrained relations that these objects must comply with. Typical constraints are given by: distance between two points, angle between two lines, belonging to a point, a tangency between two circles. On the basis of these specifications and with appropriate modeling, a constraint solver analyzes the problem and produces a construction process in the case of well-defined problems. The term well defined refers to configurations having exactly the number of constraints required for their definitions and which can be solved by this solver. After the solver proposed by Owen (1991), the domain has experienced an abundance of works using different approaches to solve a system of constraints. We quote not exhaustively (Ait-Aoudia, Jegou, & Michelucci, 1993) (Bouma, Fudos, Hoffmann, & Paige, 1995), (Latham & Middleditch, 1996), (Lamure & Michelucci, 1995), (Fudos & Hoffman, 1997), (Hoffman, Andrew, & Meera, 2001), (Michelucci & Foufou, 2004) (Owen & Power, 2007), (Ait-Aoudia & Foufou, 2010), (Cheng, Ni, & Liu, 2014), (Gao, Zhang, & Lu, 2015).

In the graph-based methods, the geometric form to be found is modeled by a non-oriented graph $G=(V, E)$ where $|V|=n$ and $|E|=m$. The geometrical elements (i.e. the set V) are represented by the nodes of the graph and the constraints (relations that must be checked by the objects) are the edges of the graph (the set E). The class of configurations solved by these methods is typically a subset of ruler and compass constructible problems with the assumption that constraint values are themselves ruler and compass constructible. An example of geometric modeling by constraints is given by the geometric form given in figure and its corresponding constraint graph is shown in figure 2 (Moussaoui & Ait-Aoudia, 2016). The graph-based denomination used in this paper will designate methods that thus model a constraint problem.

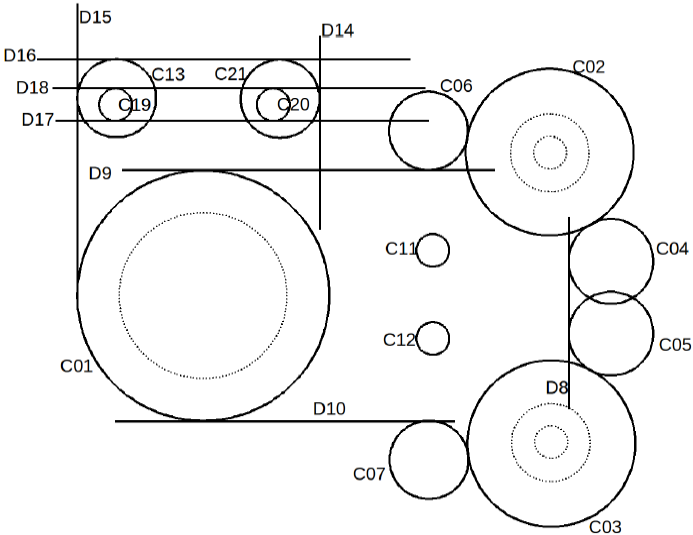

**Figure 1.** A 2D geometric design with constraints.

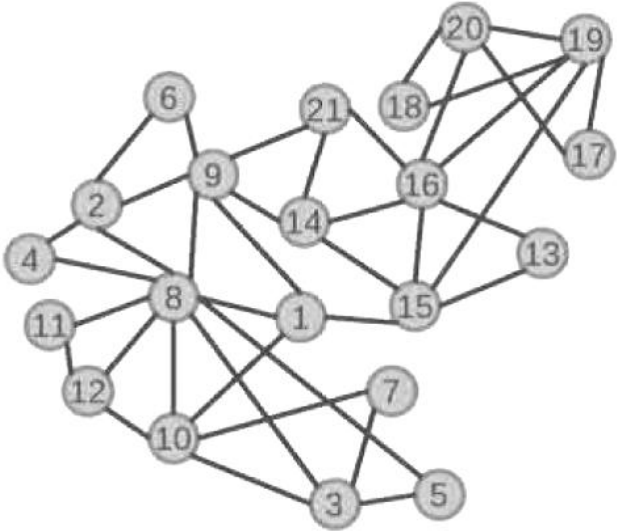

**Figure 2.** Geometric constraint graph.

Graph-based methods for solving geometric constraint problems generally operate in two phases: an analysis phase and a construction phase. These methods are often called decomposition-recombination methods. During the first phase, the constraint graph is traversed to attempt to break it down into small sub-problems. A construction sequence is derived therefrom. During the second phase, the sub-problems are solved and a combination of the solutions is carried out to give the solution of the design.

The resolution algorithms exploit the structural properties of the constraint graph to establish a resolution process. These algorithms identify in this graph the well-constrained, under-constrained and over-constrained sub-graphs. A graph is well-constrained if the number of constraints is such that it makes it possible to have finite realizations of the geometric form. A graph is under-constrained if the number of constraints is insufficient i.e. we can have an infinity of realizations of the geometric form. A graph is over-constrained if the constraints are given in excess i.e. one cannot have realization of the geometric form. A more concise definition comes from graph theory and the notion of rigid structures. This property is referred to as well-constrained graphs or structures in the domain of geometric constraints solving. We will use interchangeably well-constrained graphs and rigid structures. A characterization of rigid structures was given in 1911 by Henneberg (1911) for which any minimal rigid plane structure is obtained starting from an edge joining two vertices and adding one vertex at a time using one of the two following operations. *Operation HI*: add a new vertex *v* to *G*, then connect *v* to two chosen vertices *u* and *w* from *G* via two new edges *(v, u)* and *(v, w)*. *Operation HII*: add a new vertex *v* to *G*, chose an edge *(u, w)* and another vertex *z* from *G*, then add three edges *(v, u), (v, w)* and *(v, z)* to *G*, finally delete the edge *(u, w).* The Moser spindle shown in figure 3 is an example of a rigid structure. Henneberg construction of the Moser spindle is shown in figure 4. Well after Henneberg's work, Gerard Laman (1970) characterizes the minimally rigid plane structures composed of bars and joints by the so-called Laman graphs. Removing a bar leads to the non-rigidity or flexibility of the structure. Laman's theorem (1970), as set out below, describes these aspects.

Theorem 1.
   A graph of constraints *G=(V, E)* where *|V|=n* and *|E|=m* is structurally well constrained if and only if *m=(2\*n-3)* and *m'≤(2\*n'-3)* for any induced sub-graph *G'= (V',E ')* where *|V'| =n'* and *|E'| = m '*.

Theorem 2.
   A graph of constraints *G=(V,E)* contains a structurally over-constrained part if there is an induced sub-graph *G'=(V',E')* having more than *2\*n'-3* edges.

Theorem 3.
   A graph of constraints *G=(V,E)* is structurally under-constrained if it is not over-constrained and the number of edges is less than *2\*n-3*.

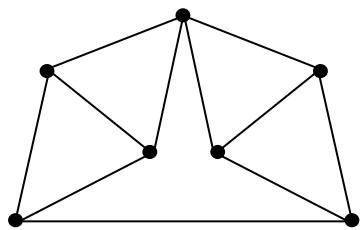

**Figure 3.** Moser spindle.

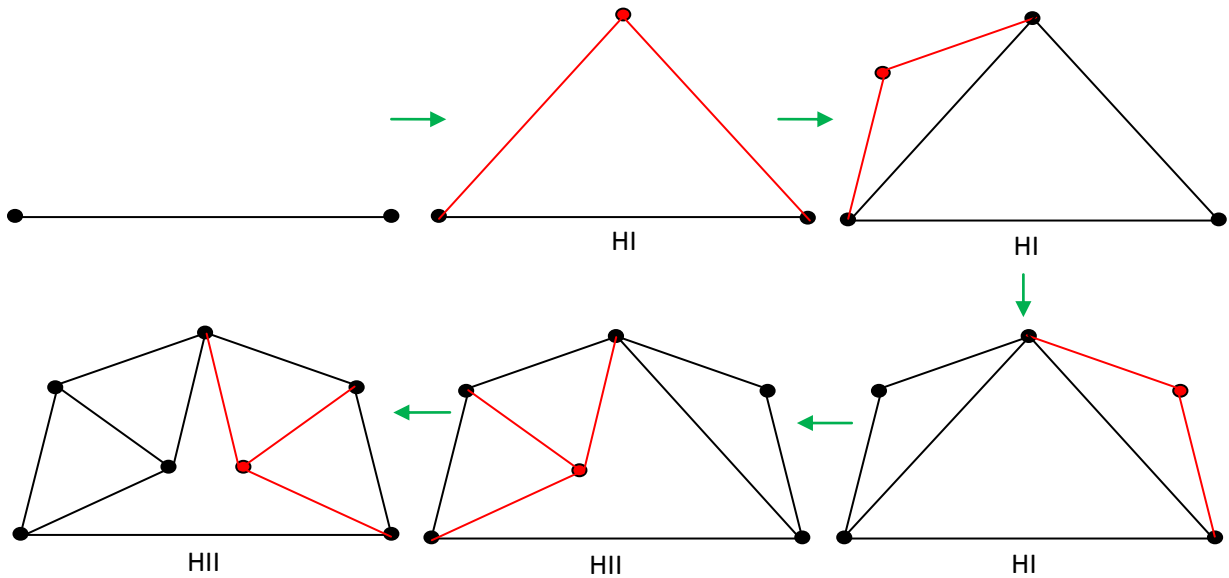

**Figure 4.** Henneberg construction of the Moser spindle.

Laman's theorem considers points in Euclidean space with two degrees of freedom and distance constraints between points. Straight lines and circles were not considered by this theorem. It should be noted that the circle of known radius has 2 degrees of freedom while the circle of unknown ray has three degrees of freedom. A generalization of this theorem to include lines and circles is considered by many authors. It should be noted that this is a rule of thumb and not a theorem.

General rule.

A graph of constraints *G=(V,E)* where *|V|=n,|E|=m, n=$n_2$+$n_3$, $n_2$ being the number of entities with 2 degrees of freedom and $n_3$ number of entities with 3 degrees of freedom,* is structurally well constrained if and only if *m=2\*$n_2$+3\*$n_3$-3* and *m'≤ 2\*$n_2$'+3\*$n_3$'-3* for any induced sub-graph *G'=(V',E')* where *|V'|=n' ,|E'|=m', n'=$n_2$'+$n_3$', $n_2$' being the number of entities with 2 degrees of freedom and $n_3$' number of entities with 3 degrees of freedom.*

Remark 1.

In some special cases, the extension of Laman's theorem can lead to incorrect diagnostics. A typical example of such a case (Ait-Aoudia & Foufou, 2010) is given by a constrained triangle with three angles. In figure 4.a, a triangle is defined with three angular constraints (where α+β+γ=180°). This triangle is geometrically under-constrained but its constraint graph, shown in Figure 4.b, is well constrained. It is pointed out that this case will still be detected during the construction phase.

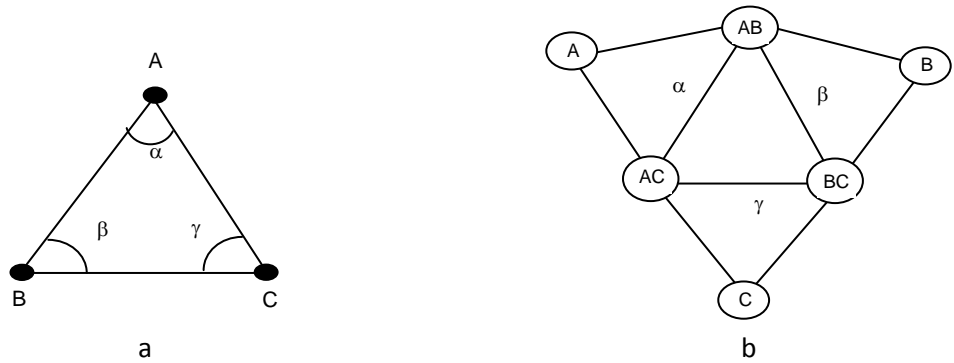

Figure 5. An under-constrained design (a), with the corresponding constraint graph (b).

Remark 2.

The graph-based constructive approach uses only the structural properties of the constraint graph and does not take into account the numerical information. A constraint graph can be structurally well-constrained but numerically under-constrained or without solutions. An example of a numerically under-constrained geometric sketch is shown in Figure 5.a (values indicate distance measurements). Its corresponding constraint graph shown in Figure 5.b is structurally well constrained because it satisfies the Laman condition (1970). Constraint values can lead to graphs without a numerical solution. The intersection of two circles, for the construction of a new point, may be empty. These cases will nevertheless be detected during the construction phase.

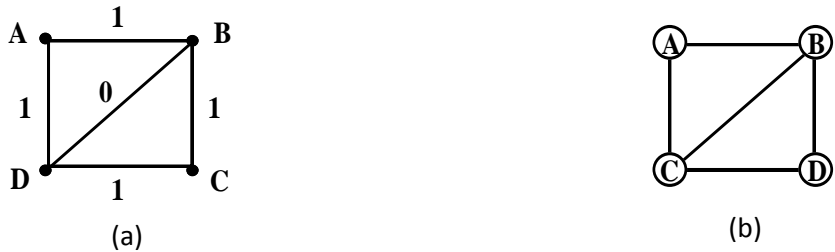

Figure 6. A numerically under-constrained design (a) with the corresponding constraint graph (b)

## II. RULER AND COMPASS CONSTRUCTIONS

Graph-based solvers are often linked by their developers to ruler and compass constructions. Indeed these constructions are at the heart of various fields of application. Historically these constructions are central in Euclid's elements. Moreover, Euclidean geometry is often considered as the geometry of the "ruler and compass". Straight lines and circles were considered by Greek mathematicians as ideal figures. Several problems in geometry were set out to see if they were constructible with ruler and compass. Some problems have stood two millennia to have a negative answer like squaring the circle, the duplication of the cube, the regular polygon with 7 sides or the trisection of an angle.

The ruler and compass constructions can be obtained by the following method: starting from a set of base points denoted B, two elementary constructions are used to construct new points. The two basic constructions are:

– Draw with the ruler, the line passing through two points of B

- Draw with a compass the circle having a point of B as its center and the distance between two points of B as radius.

The intersection points of the newly drawn lines and circles are added to the points base B. This process can be repeated at infinity. This construction method is formalized by the definition of Carrega (2001) given hereinafter:

Definition :
Let P be a Euclidean plane and B a finite subset of P having at least two elements. The elements of B are called base points.

- A point M of P is said to be ruler and compass constructible from B if there exists a finite sequence of points of P, $M_1$, $M_2$, … $M_n$ ending with M such that for all i, 1≤i ≤n, Mi is a point of intersection either of:
  - two straight lines
  - a line and a circle
  - two circles

  These straight lines and circles being obtained using the set $E_i$=B ∪ { $M_1$, $M_2$, … $M_n$} as follows:
  - each line passes through two distinct points of $E_i$
  - each circle is centered at a point of $E_i$ and has as its radius the distance between two points of $E_i$.

- A line passing through two constructible points is said to be constructible.
- A circle centered at a constructible point and having as radius the distance between two constructible points is constructible.

Drawing new points with the ruler and to the compass amounts to solving linear or degree 2 algebraic equations. Indeed, intersecting two straight lines amounts to solving a linear equation, and intersecting a line and a circle or intersecting two circles amounts to solving an equation of degree 2.

To characterize algebraically the constructions with the ruler and the compass, we consider the Euclidean plane P, with origin the point O of coordinates (0,0) and a second point I (1,0) on the abscissa axis. These two points constitute the starting point for any construction with the ruler and the compass. Construction problems are reduced to algebraic problems. We recall below some definitions and theorems.

**Definition 1.** A real number x ∈ R is said to be constructible if the point of coordinates (x,0) on the abscissa axis can be constructed with ruler and compass.

**Definition 2.** A point M ∈ P of coordinates (x,y) is constructible with rule and compass if and only if its two coordinates x ∈ R and y ∈ R are real constructible numbers.

**Wantzel's theorem.** A real number x is constructible if and only if there exist quadratic extensions (K0 = Q) ⊂ K1 ⊂ … ⊂ Kc such that x ∈ Kc. Each of the field extensions is quadratic i.e. [Ki + 1: Ki] = 2. In other words, each extension is a quadratic extension of the preceding one: Ki+1 = Ki($\sqrt{di}$) for some di ∈ Ki.

**Corollary of Wantzel's theorem.** Any real constructible number is an algebraic number on Q whose algebraic degree is of the form $2^n$, $n \geq 0$.

## III. RESOLUTION

The general principle of resolution of the graph-based methods is to find a process of incremental construction of the geometric figure using at each stage a drawing with the rule and the compass. Various techniques have been proposed to achieve this goal. These methods for solving constrained geometric problems can be divided into two main categories, namely top-down methods and bottom-up methods.

Top-down methods proceed by a recursive decomposition of the initial constraint graph until we arrive at so-called elementary (soluble in one step) graphs whose resolution is trivial. The actual construction process proceeds in the opposite direction to the decomposition process.

Bottom-up methods identify the elementary graphs and proceed by successive groupings by elementary rules until obtaining the initial graph. By fixing for example two geometric entities in the 2D space, one proceeds by successive addition of entities attached to the figure fixed by a number of edges equivalent to their degree of freedom.

For example, a constrained design and its corresponding constraint graph is shown in figure 7. Figure 8 illustrates the decomposition of the latter constraint graph by the method described by Ait-Aoudia et al. (1999). Triangulated graphs called clusters are obtained in the first phase (Figure 7a). Virtual edges deduced from the clusters are added in order to assemble the figure solution (figure 7.b).

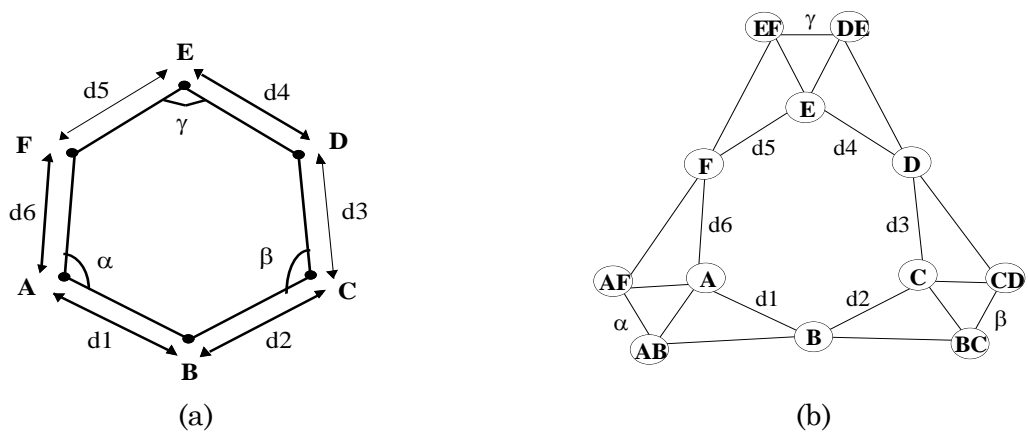

**Figure 7.** Une figure et son graphe de contraintes.

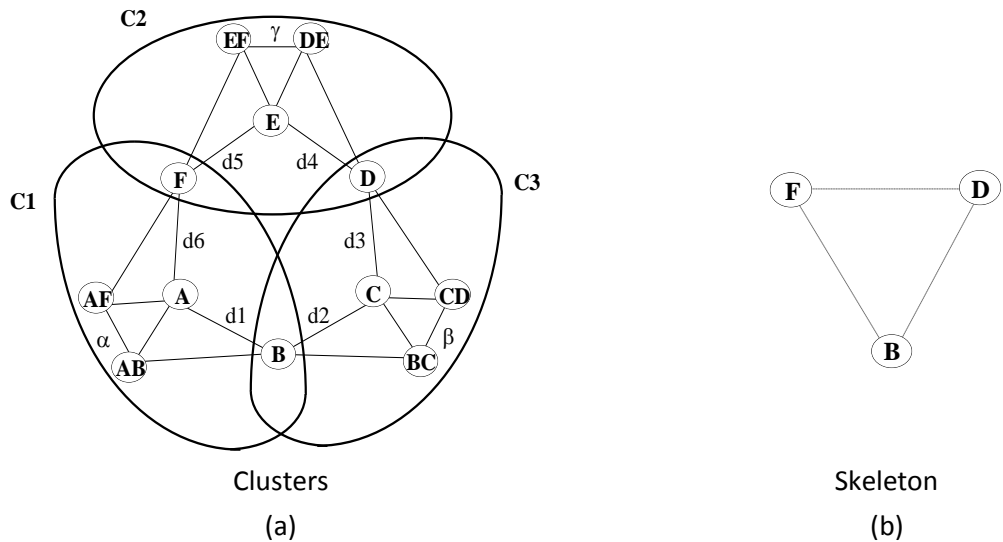

Clusters
(a)

Skeleton
(b)

**Figure 8.** Résolution par une méthode ascendante.

## IV. SOLUBLE GRAPHS AND CONSTRUCTIONS WITH RULER AND COMPASS

The graph-based methods generally solve well geometric designs constructible with ruler and compass, but stumble on certain configurations whose constraint graph cannot be decomposed. We must, however, distinguish between decomposable constraints graphs and constructions which can be drawn with the rule and with the compass, two concepts often mistakenly assimilated. Indeed, some authors hastily assert, like Owen (1991) and Lee et al. (2003), that any configuration that cannot be solved by their respective solvers cannot be drawn with ruler and compass. The following example (Ait-Aoudia, 1994) gives a counter-example to this assertion.

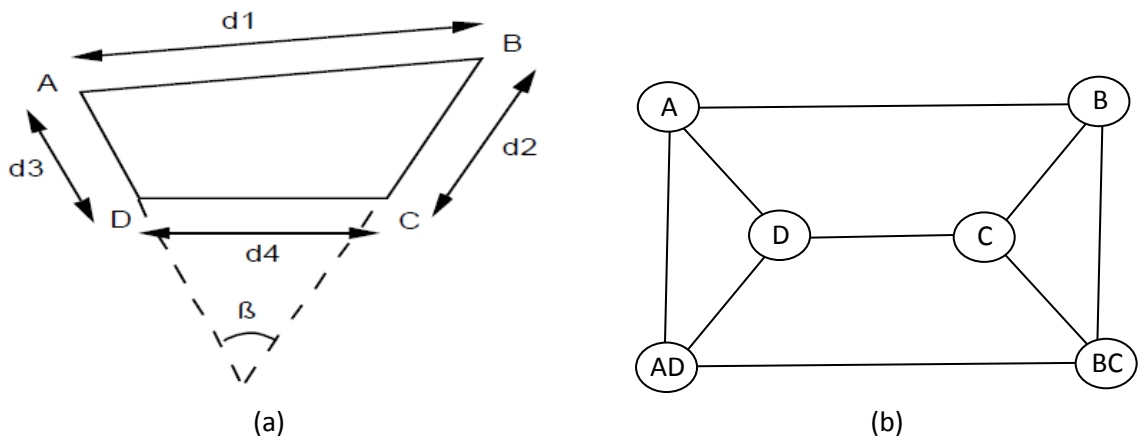

(a)            (b)

**Figure 9.** Geometric design and corresponding constraint graph.

Let the quadrilateral ABCD given by its four distances and an angle between two opposite sides AD and BC (FIG. 7a). The corresponding constraint graph is given in Figure 7.b. This graph of constraints is tri-connected and therefore not decomposable by the algorithm of Owen or Lee et al., whereas the geometric figure is perfectly constructible with ruler and the compass. To construct figure 7.a using ruler and compass (figure 8a), proceed as follows: the points A and D are fixed, the segment AE is drawn such that the angle (AE , AD) is equal to ß (the given angle between AD and BC) and the distance AE is equal to the distance CB. The segment AE is parallel to the segment CB. The point C is then obtained by the intersection of the two circles of centers D and E and of respective radii d4 and d1. The point B is finally obtained by the intersection of the two circles of centers C and A of respective radii d2 and d1.

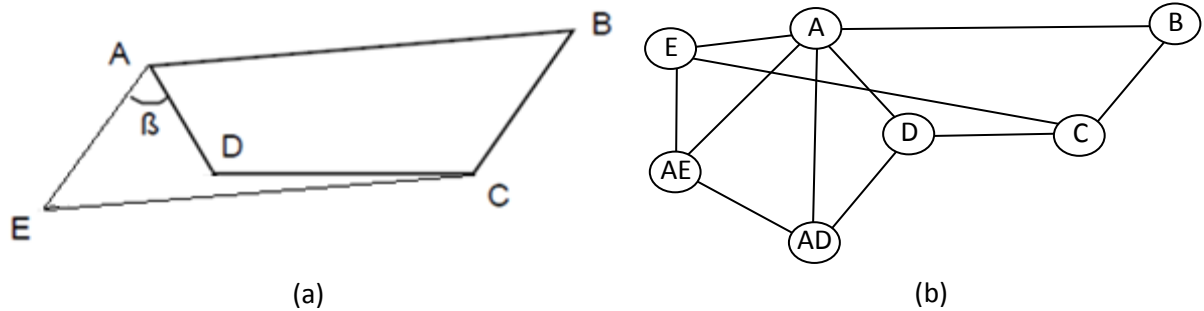

(a)                                                                                              (b)

**Figure 10.**   Ruler and compass construction and corresponding constraint graph.

This construction uses an extra point E to get the final result, something that decomposition algorithms do not take into account in their construction process. These algorithms ignore intermediate constructs. If one constructs the graph of the constraints corresponding to this method of construction with the point E, this graph becomes perfectly decomposable as shown in figure 8.b.

To further illustrate this remark, we take two famous examples constructible with ruler and compass and whose corresponding constraint graphs are not resolved by the aforementioned methods.

The first example is the Cramer-Castillon problem illustrated in figure 9 and whose statement is "given a circle $\Gamma$ of center O and of radius R and three points A, B and C, build, using only the ruler and the compass, a triangle MNP inscribed in the circle and whose sides pass respectively by the points A, B and C ". The corresponding constraint graph is given in Figure 10.

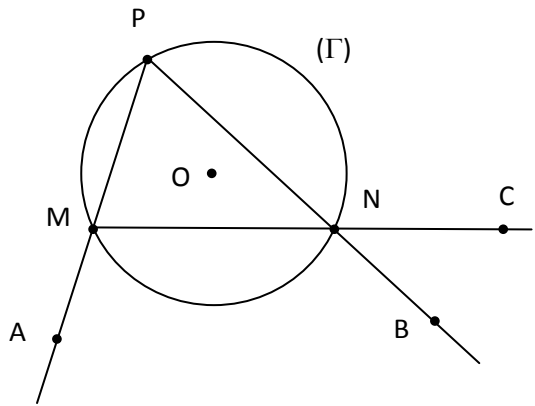

**Figure 11.**   Cramer-Castillon problem.

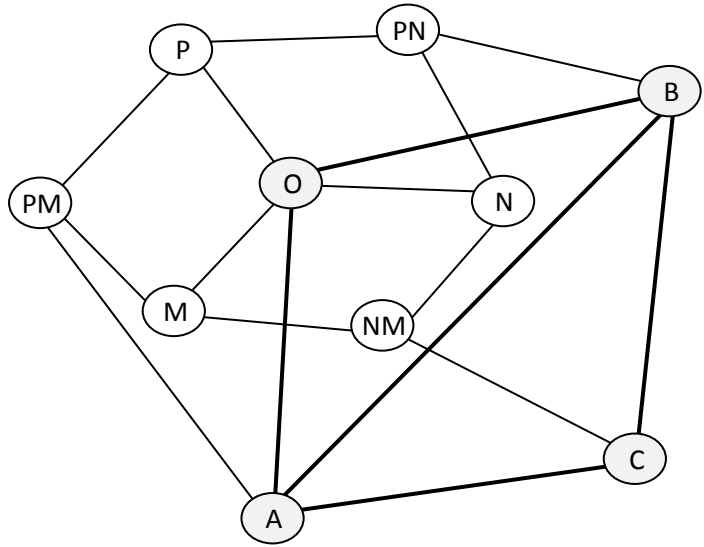

**Figure 12.** Constraint graph of Cramer-Castillon problem.

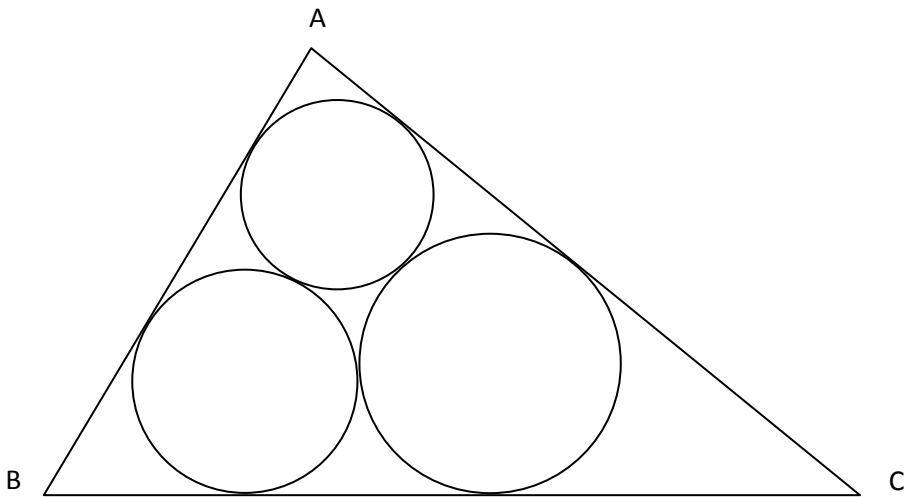

**Figure 13.** Malfatti problem.

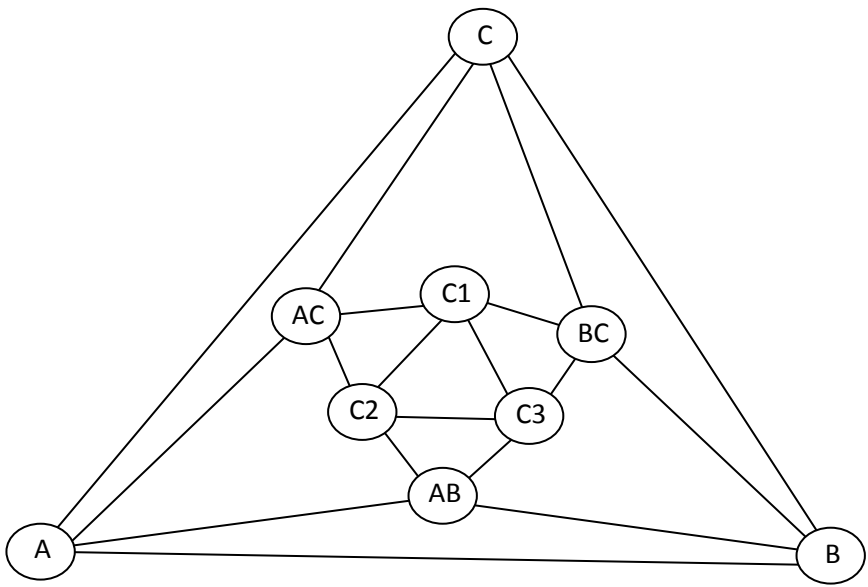

**Figure 14.** Constraint graph of Malfatti problem.

The second example is the Malfatti problem illustrated in Figure 11 and whose statement is "Given a triangle ABC, construct with the ruler and compass three circles C1, C2 and C3 inscribed in the angles A, B and C of the triangle and such that each of them is tangent to the other two ".

Constraint solvers cannot solve these graphs and declare them non-constructible to rule and compass. This false observation comes from the fact that the non-trivial intermediate constructs indispensable to the rule and compass construction process are not materialized in the constraint graph.

## V. REDUCIBILITY OF CONSTRAINT GRAPHS

This clarification being made, we will distinguish two classes of configurations that cannot be solved with graph-based methods:

i. Partially reducible graphs: these configurations can be solved partially but not entirely. In this type of graph decomposable sub-graphs can be detected, but global recombination is impossible with basic construction rules. An example of such a graph is given by the following figure. Vertices represent points and edges are distance constraints. We can detect both triangles but the overall resolution is impossible.

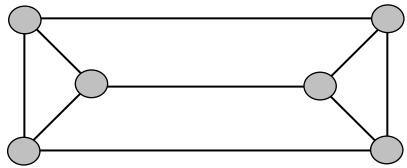

**Figure 15.** Partially reducible graph.

ii. Irreducible graphs: these configurations are totally non decomposable. We will designate them by N-IR-Graph, N denoting the number of geometric entities or vertices of the graph. In this type of graph no decomposition is possible. A 6-IR-Graph, with vertices representing points and edges representing distance constraints, is given in FIG. 14. Irreducible graphs for N ranging from 6 to 100 and generated by Moussaoui et al. (2016) are given in (Moussaoui & Ait-Aoudia, 2016) .

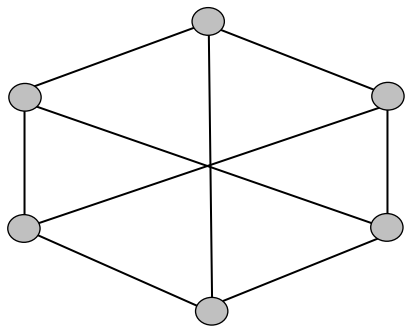

**Figure 16.** 6-IR-Graphe

VI. CONCLUSION

The configurations solved by the graph-based methods are figures constructible with the ruler and the compass. These methods have the advantage of being able to provide a geometric explanation to the user during the resolution phase. On the other hand, they fail on configurations constructible with the ruler and the compass and whose intermediate stages of construction are not materialized in the constraint graph. The same goes for irreducible graphs where no decomposition is possible. It is then necessary to have an algorithm able to explain the failure of the resolution and to switch on a numerical resolution (Ait-Aoudia & Mana, 2004). Constructability tests on the rule and the compass of certain configurations are given by Gao and Chou (1998) and Schreck and Mathis (2016).